\documentclass[aps,prl,reprint,groupedaddress,showpacs]{revtex4-1}

\usepackage{graphicx}
\usepackage{natbib}

\begin{document}


\title{Field Tuning of Ferromagnetic Domain Walls on Elastically Coupled Ferroelectric Domain Boundaries}


\author{K\'{e}vin J. A. Franke, Tuomas H. E. Lahtinen, Sebastiaan van Dijken}
\email[]{sebastiaan.van.dijken@aalto.fi}
\affiliation{NanoSpin, Department of Applied Physics, Aalto University School of Science, P.O. Box 15100, FI-00076 Aalto, Finland.}


\date{\today}

\begin{abstract}
We report on the evolution of ferromagnetic domain walls during magnetization reversal in elastically coupled ferromagnetic-ferroelectric heterostructures. Using optical polarization microscopy and micromagnetic simulations, we demonstrate that the spin rotation and width of ferromagnetic domain walls can be accurately controlled by the strength of the applied magnetic field if the ferromagnetic walls are pinned onto 90$^\circ$ ferroelectric domain boundaries. Moreover, reversible switching between magnetically charged and uncharged domain walls is initiated by magnetic field rotation. Switching between both wall types reverses the wall chirality and abruptly changes the width of the ferromagnetic domain walls by up to $1000\%$.        
\end{abstract}


\maketitle

Ferromagnetic and ferroelectric domain walls have been studied for many decades and their static and dynamic properties are still a topic of intense investigation. This ongoing interest is stimulated by the discovery and potential application of current-induced magnetic domain wall motion \cite{1978JAP....49.2156B,*1984JAP....55.1954B,2008Sci...320..190P}, by advances in nanofabrication and high-resolution characterization techniques \cite{Hopster,Alexe,2008NatMa...7...57J}, and by the use of ferromagnetic and ferroelectric materials in practical devices including non-volatile memories, sensors and actuators, and microwave components \cite{2007NatMa...6..813C,2006JAP...100e1606S}. In ferromagnets, the domain wall structure is determined by a competition between exchange, anisotropy, and magnetostatic energies. As a result, the domain wall width varies from a few nanometer in hard or geometrically constrained magnets, to several tens of nanometer in bulk Co and Fe, and to hundreds of nanometer in low anisotropy materials or thin films with uniform N\'{e}el walls \cite{HubSchaf}. In ferroelectric materials, domain walls are more abrupt. The width of 180$^\circ$ electrostatic walls in tetragonal PbTiO$_3$, PbZr$_{0.2}$Ti$_{0.8}$O$_3$, and BaTiO$_3$, for example, is only one to two lattice constants \cite{1999ApPhL..75.2830P,2002PhRvB..65j4111M,2008NatMa...7...57J,2010PhRvB..81n4125M}. Ferroelectric domain walls that are predominantly determined by elastic strains are wider, yet their width is still small compared to their ferromagnetic counterparts. Experimental and theoretical results on 90$^\circ$ ferroelastic walls in BaTiO$_3$ indicate that most of the polarization rotates within 2 - 5 nm \cite{1992ApPhL..60..784Z,2006PhRvB..74j4104H,2006ApPhL..89r2903Z}.

While the properties of ferromagnetic and ferroelectric domain walls in single ferroic systems are meanwhile understood, rigorous results on the structure and manipulation of coupled ferromagnetic-ferroelectric walls are scarce. Given the different nature and scaling of ferromagnetic and ferroelectric walls, strongly interacting systems are expected to display a rich variety of static and dynamic phenomena. Moreover, the realization of interferroic domain wall coupling opens up additional degrees of freedom to tune internal wall structures and to control domain wall motion.  

In this letter, we report on the properties of N\'{e}el-type ferromagnetic domain walls on top of ferroelectric substrates with 90$^\circ$ ferroelastic stripe domains. The experimental results obtained by optical polarization microscopy are complemented by micromagnetic simulations. We demonstrate that strain transfer from alternating ferroelastic domains strongly couples the ferromagnetic domain walls to the underlying ferroelectric boundaries. As a result, the domain wall width and total spin rotation within the walls can accurately be tuned by the magnetic field strength. Moreover, rotation of the field direction induces abrupt and reversible switching between broad charged and narrow uncharged ferromagnetic domain walls, a feature that could be used in practical applications. 

\begin{figure}
\includegraphics{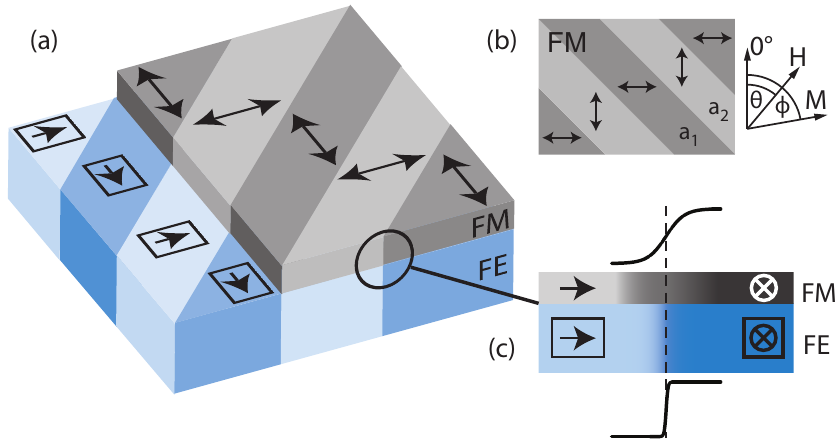}
\caption{\label{fig1} (a) Schematic illustration of the ferroelectric and ferromagnetic domain pattern in the Co$_{60}$Fe$_{40}$ - BaTiO$_3$ heterostructures. (b) Definition of the magnetization angles ($\phi$) in the a$_1$ and a$_2$ domains and the direction of the applied magnetic field ($\theta$). (c) Relatively broad ferromagnetic domain walls are pinned onto narrow 90$^\circ$ ferroelectric boundaries. The arrows and double-headed arrows indicate the orientation of the ferroelectric polarization and the strain-induced uniaxial magnetic easy axis in the a$_1$ and a$_2$ domains.}
\end{figure}

In the experiments, thin ferromagnetic Co$_{60}$Fe$_{40}$ films combining large magnetostriction and small magnetocrystalline anisotropy \cite{1960JAP....31S.157H} were grown onto ferroelectric BaTiO$_3$ single-crystal substrates with a regular in-plane 90$^\circ$ ferroelastic domain pattern using electron-beam evaporation. At room temperature, the direction of ferroelectric polarization in BaTiO$_3$ coincides with a tetragonal lattice elongation of 1.1\% inducing 90$^\circ$ rotations of the magnetoelastic anisotropy axis in the Co$_{60}$Fe$_{40}$ film via interface strain transfer and inverse magnetostriction. Due to the absence of other significant magnetic anisotropy contributions, full transfer of the ferroelectric domain pattern to the ferromagnetic film is obtained \cite{ADMA:ADMA201100426,2011ITM....47.3768L}. In this configuration, ferromagnetic domain walls are pinned onto narrow ferroelastic boundaries and magnetization reversal in neighboring domains proceeds largely independently. The ferromagnetic-ferroelectric system under study is schematically illustrated in Fig. \ref{fig1}. Optical polarization microscopy techniques were used to image the ferroelectric and ferromagnetic domain structures. Contrast from both ferroic domains was obtained by selecting ferromagnetic thin films with a thickness that renders them semi-transparent for visible light (typically 10 - 20 nm) and by switching between birefringent and magneto-optical Kerr effect contrast for imaging of BaTiO$_3$ and Co$_{60}$Fe$_{40}$ domains, respectively.

\begin{figure}
\includegraphics{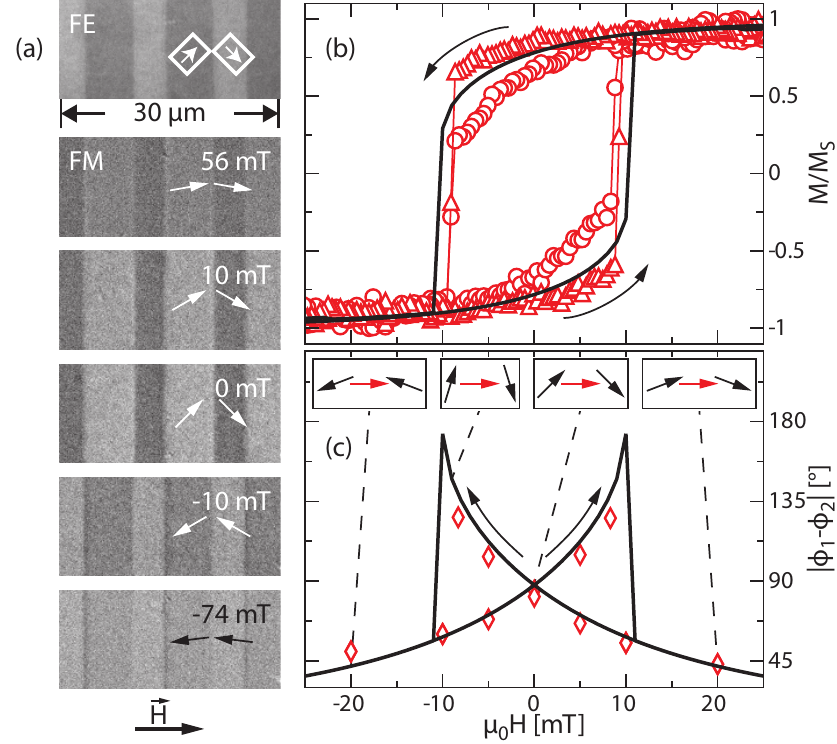}
\caption{\label{fig2} (a) Polarization microscopy images of the ferroelectric domain pattern in the BaTiO$_3$ substrate (FE) and the ferromagnetic domains in the overlaying Co$_{60}$Fe$_{40}$ film (FM) during several stages of the magnetization reversal process. The field is applied perpendicular to the domain walls and the axis of magneto-optical Kerr effect contrast is parallel to the walls. (b) Experimental hysteresis curves of two neighboring a$_1$(circles) and a$_2$(triangles) domains and the calculated result from micromagnetic simulations (line). (c) Experimentally determined (diamonds) and simulated (line) spin rotation within the ferromagnetic domain walls as a function of applied magnetic field. The black arrows in (c) indicate the magnetization direction in neighboring domains and the red arrows illustrate the spin orientation in the center of the domain walls during a field sweep from 20 mT to -20 mT.}
\end{figure}

Figure \ref{fig2}(a) shows domain images of a Co$_{60}$Fe$_{40}$ film at different stages of the magnetization reversal process for an applied magnetic field perpendicular to the domain walls. The ferroelectric stripe pattern of the BaTiO$_3$ substrate (FE) is magnetically reproduced in the Co$_{60}$Fe$_{40}$ film (other images). The magnetic pattern is robust and it is only erased by large magnetic fields that fully saturate the film magnetization. With decreasing field strength, the magnetization of neighboring stripe domains rotate in opposite directions towards their respective easy magnetoelastic anisotropy axes while the walls separating the domains are fully immobilized by elastic coupling to the underlying ferroelectric boundaries. This reversal mechanism contrasts with the more conventional micromagnetic response where the Zeeman energy provided by an external magnetic field is accommodated by lateral domain wall motion, while intrinsic wall properties such as width and spin rotation remain largely the same. In our ferromagnetic-ferroelectric heterostructures, 90$^\circ$ rotations of the magnetoelastic anisotropy axis result in strong domain wall pinning and lateral modulations of magnetization reversal and, as a result, the structure of the ferromagnetic domain walls changes continuously as a function of applied magnetic field. An example is shown in Fig. \ref{fig2}(c). Here, the total spin rotation within the ferromagnetic walls is plotted versus the field strength perpendicular to the domain walls. The experimental data (diamonds) were obtained by measuring local magnetic hysteresis curves on two neighboring  a$_1$ and a$_2$ stripe domains (Fig. \ref{fig2}(b)). From such measurements, $\phi_{1}$ and $\phi_{2}$ were determined by considering coherent magnetization reversal prior to abrupt magnetic switching in both domains as clearly indicated by magneto-optical Kerr effect images (Fig. \ref{fig2}(a)). Obviously, clockwise and anti-clockwise magnetization rotation in neighboring domains increases the spin rotation within the ferromagnetic domain walls from zero in saturation, to 90$^\circ$ in remanence, and finally to nearly 180$^\circ$ just before abrupt magnetic switching. After the switching event, the angle between the magnetization of neighboring domains is about 60$^\circ$ and this reduces gradually with a further increase of the magnetic field. A closer inspection of the magnetization reversal process learns that the actual spin rotation within the walls is  360$^\circ-|\phi_{1}-\phi_{2}|$ after switching because the magnetization in the center of the wall does not rotate during this process. The formation and properties of these N\'{e}el-type double walls, which have also been observed in other systems \cite{HubSchaf,1999JAP....85.5160C,2000PhRvB..61..580T,2000ApPhL..76..754P,2006JAP...100d3918L}, will be discussed in more detail elsewhere.  

To further analyze the field tunability of ferromagnetic domain walls that are elastically pinned onto 90$^\circ$ ferroelectric boundaries, micromagnetic simulations were conducted using object oriented micromagnetic framework (OOMMF) software \cite{NIST}. In the micromagnetic simulations it is assumed that the structure of the ferroelectric boundaries does not change during magnetization reversal in the ferromagnetic film. This assumption is supported by an estimation of the maximum strain that can be transferred from the Co$_{60}$Fe$_{40}$ film to the BaTiO$_3$ substrate via magnetostriction. For isotropic ferromagnetic films, the anisotropic magnetostrictive strain relative to the direction of magnetization is given by $\epsilon=3/2\lambda_{s}($cos$^{2}(\phi)-1/3)$ 
\cite{Handley}, where $\lambda_{s}$ is the magnetostriction constant ($\approx$ $6.8\times10^{-5}$ for Co$_{60}$Fe$_{40}$ \cite{1960JAP....31S.157H}). Hence, the maximum anisotropic strain for 90$^\circ$ magnetization rotation equals 3/2$\lambda$ = $0.01\%$. This change in lattice elongation is two orders of magnitude smaller than the anisotropic strain of the BaTiO$_3$ substrate ($1.1\%$ at 90$^\circ$ boundaries) and therefore magnetostriction does not provide enough elastic energy to significantly alter the ferroelectric domain walls. Asymmetric strain transfer between the two ferroic materials is also confirmed by experiments. While electric-field driven modifications of the ferroelectric domain pattern directly alter the ferromagnetic microstructure \cite{ADMA:ADMA201100426,2011ITM....47.3768L}, the ferroelectric stripe pattern does not change upon the application of a magnetic field in any of our measurements. 

In the micromagnetic simulations, an uniaxial magnetoelastic anisotropy of $K_{me}=1.7\times10^4$ J/m$^3$ was used. This value was experimentally determined from the slope of hard-axis magnetization curves on single stripe domains. Other parameters include a saturation magnetization of $M_s=1.7\times10^6$ A/m, a uniform exchange constant of $K_{ex}=2.1\times10^{-11}$ J/m, a film thickness of 10 nm and an in-plane unit mesh of $2\times2$ nm. The total area of the simulations contained two 5 $\mu$m wide stripe domains and two-dimensional boundary conditions were applied to avoid artificial edge effects \cite{wang_two-dimensional_2010}.         

\begin{figure}
\includegraphics{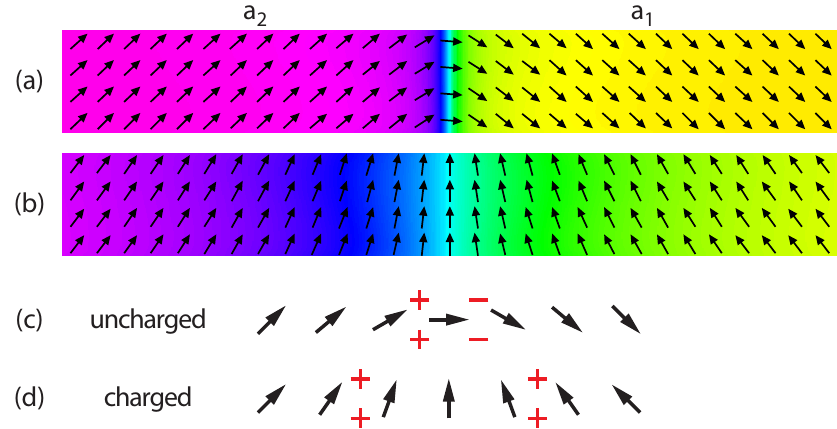}
\caption{\label{fig3}Remanent micromagnetic structure for $K_{me}=1.7\times10^4$J/m$^3$ after the application of a magnetic field (a) perpendicular and (b) parallel to the ferromagnetic domain wall. (c), (d) Schematic illustration of a narrow uncharged and a broad charged ferromagnetic domain wall.}   
\end{figure}

\begin{figure}
\includegraphics{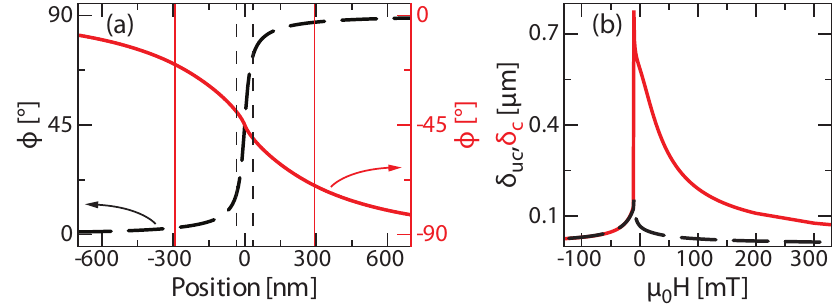}
\caption{\label{fig4} (a) Wall profile of the uncharged (black dashed line) and charged (red solid line) walls in magnetic remanence. Vertical lines indicate the calculated wall width using $\delta=\int_{-\infty}^{+\infty}$cos$(\phi)^2$d$x$. (b) Tuning of the uncharged and charged domain wall width as a function of magnetic field strength.}   
\end{figure}

The micromagnetic simulations largely reproduce the measured hysteresis curves on single stripe domains (Fig. \ref{fig2}(b)) and the variation of total spin rotation within the ferromagnetic domain walls with applied magnetic field (Fig. \ref{fig2}(c)). The difference in the experimental data is most likely due to a variation in stripe width (see domain images of Fig. \ref{fig2}(a)), whereas complete symmetry in the simulations produces identical hysteresis curves for both domains. Figures  \ref{fig3}(a) and \ref{fig3}(b) show the magnetic microstructures in remanence after applying a large magnetic field perpendicular and parallel to the domain walls. In both cases, the total spin rotation within the ferromagnetic wall rotates by 90$^\circ$ and the magnetization in the center of the wall aligns with the field direction. The width of the domain walls, however, is very different for these field geometries as illustrated by the wall profiles in Fig. \ref{fig4}(a). For fields perpendicular to the walls, the magnetization vectors of neighboring domains do not create any net magnetic charges (Fig. \ref{fig3}(c)). In this case, the criterium for uncharged walls, $(\mathbf{M_{a1}}-\mathbf{M_{a2}})\mathbf{\cdot}\mathbf{\hat{n}}=0$ where $\mathbf{\hat{n}}$ indicates the wall normal, is satisfied. Using $\delta=\int_{-\infty}^{+\infty}$cos$(\phi)^2$d$x$ with $x$ along $\mathbf{\hat{n}}$ to calculate the domain wall width from the micromagnetic simulations \cite{HubSchaf}, we find $\delta_{uc}=68$ nm. When the field is parallel to the walls, the magnetization aligns in a head-to-head configuration and this charges the walls (Fig. \ref{fig3}(d)). In bulk magnetic samples, the magnetostatic charging energy is large and, hence, the formation of charged domain walls is extremely rare. In thin films, charged walls are more common since their magnetostatic energy reduces with decreasing film thickness. In our ferromagnetic-ferroelectric heterostructures, 90$^\circ$ rotations of the magnetoelastic anisotropy axis impose the formation of an array of charged walls when the field component parallel to the walls exceeds the perpendicular component. The width of the magnetically charged wall in Fig. \ref{fig3}(b) is  $\delta_{c}=587$ nm, which is more than $8$ times larger than the width of the uncharged wall in the same system. 

Besides the ability to select the type of N\'{e}el wall in ferromagnetic-ferroelectric heterostructures, the width of both domain walls can continuously be tuned by variation of the magnetic field strength. This is illustrated in Fig. \ref{fig4}(b). For large fields, the angle between the magnetization of neighboring domains is small and this results in narrow domain walls. As the strength of the field is reduced, the width of the uncharged and charged walls increase. Finally, after abrupt magnetic switching, narrow double walls form and their width gradually decreases with increasing field. The domain wall width thus varies continuously during magnetization reversal from zero at the onset of coherent spin rotation to, depending on the type of wall, several tens or hundreds of nanometer.

\begin{figure}
\includegraphics{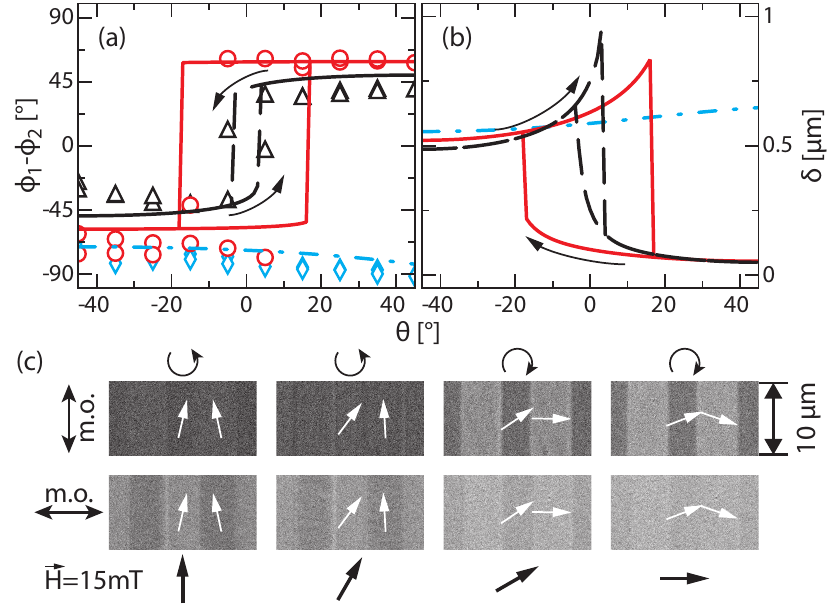}
\caption{\label{fig5} (a) Dependence of the total spin rotation $\phi_{1}-\phi_{2}$ within the ferromagnetic domain walls on magnetic field angle at constant field strength as determined by optical polarization microscopy experiments (symbols) and micromagnetic simulations (lines). (b) Corresponding change in domain wall width during magnetic field rotation. In (a) and (b) the magnitude of the applied field is $5$ mT (diamonds and dash-dotted line), $10$ mT (circles and solid line) and $15$ mT (triangles and dashed line). (c) Magneto-optical Kerr effect images of the ferromagnetic domain pattern in the Co$_{60}$Fe$_{40}$ film during magnetic field rotation from -45$^\circ$ to 45$^\circ$. The field strength was 15 mT and the images were recorded with magneto-optical (m.o.) contrast parallel (upper row) and perpendicular (lower row) to the walls. The arrows in the images indicate the magnetization direction in neighboring domains and the circular arrows above the images illustrate the chirality of spin rotation within the ferromagnetic domain walls.}
\end{figure}

Hysteretic switching between broad charged and narrow uncharged ferromagnetic domain walls occurs when the applied field direction is rotated at constant and sufficiently large field strength. Figure \ref{fig5} shows experimental and simulation results for a continuous rotation of the external magnetic field from -45$^\circ$ (parallel to the walls) to 45$^\circ$ (perpendicular to the walls) and back. If the applied magnetic field exceeds the critical field strength ($H_{c}$) for magnetic switching in one of the domains, abrupt switching from one domain wall type to the other and vice versa is triggered at well-defined field angles. These instant wall transformations are characterized by a reversal of the wall chirality from $\phi_{1}-\phi_{2}<0$ (anti-clockwise spin rotation) to $\phi_{1}-\phi_{2}>0$ (clockwise spin rotation) and an anomalous change in domain wall width. The experimentally determined evolution of the spin rotation ($\phi_{1}-\phi_{2}$), which was extracted from polarization microscopy images with different axes of magneto-optical Kerr effect contrast (Fig. \ref{fig5}(c)), agrees well with the micromagnetic simulations for all field strengths. The concurrent change in domain wall width is significant. As an example, for increasing field angle and $H=10$ mT (solid line in Fig. \ref{fig5}(b)), the ferromagnetic domain wall width abruptly drops from $\delta=846$ nm to $\delta=70$ nm at a field angle of $\theta=17^\circ$, indicating a relative effect of more than $1000\%$. Rotation of the magnetic field in the oposite direction first increases the wall width via a gradual increase of the magnetic charges on the wall, which is followed by abrupt switching to a broad domain wall at a negative field angle. For field strengths $H<H_{c}$ no switching occurs and both the spin rotation and domain wall width vary only weakly with applied field angle (diamonds and dash-dotted lines in Fig. \ref{fig5}).    
  
In summary, we have demonstrated the ability to accurately control the intrinsic properties of ferromagnetic domain walls by elastic coupling to narrow 90$^\circ$ ferroelectric domain boundaries. Both the total spin rotation within the ferromagnetic walls and the width of the walls can be tuned by varying the magnetic field strength. Moreover, reversible switching between broad charged and narrow uncharged walls with opposite chirality can be induced by magnetic field rotation. This high degree of tunability opens the door to new magnetic devices wherein domain walls are utilized as functional and controllable elements. For example, abrupt switching between two domain wall states could be used to store information or function as a magnetic switch in magnonic crystals, and tunable ferromagnetic domain walls could form the basis of tunable microwave resonators.      

This work is supported by the Academy of Finland under contract no. 127731. K.J.A.F. acknowledges support from the Finnish Doctoral Program in Computational Sciences. T.H.E.L. is supported by the National Doctoral Program in Materials Physics. 


%

\end{document}